\documentclass{rnaastex}

\usepackage{graphics,graphicx,url,verbatim,xspace}
\usepackage{glossaries}
\usepackage{comment}
\usepackage[frozencache,cachedir=.]{minted}

\definecolor{black}{rgb}{0,0,0}
\definecolor{red}{rgb}{1.0,0,0}
\newcommand{\UCB}{Department of Astronomy,  University of California, Berkeley, Berkeley, CA 94720}

\newcommand{\CIRA}{International Centre for Radio Astronomy Research, Curtin University, Bentley WA 6102, Australia}

\begin{document}

\title{Global Sky Models can Improve Flux Estimates in Pulsar and FRB Studies}


\correspondingauthor{Danny C.\ Price}
\email{danny.price@curtin.edu.au}

\author[0000-0003-2783-1608]{Danny C.\ Price}
\affiliation{\CIRA}
\affiliation{\UCB}

\begin{abstract}
It is commonplace in pulsar and fast radio burst (FRB) literature to estimate sky temperature by frequency-scaling of the \citet{Haslam:1982} 408 MHz map. I suggest that this practice should stop, in favor of using readily-available global sky models of diffuse foregrounds. This practical change will improve accuracy of pulse flux estimates.

\end{abstract}


\section{Introduction}\label{sec:intro}

In searches for fast radio bursts (FRBs) and single pulses from pulsars, it is common for the peak flux to be estimated based on the signal-to-noise (S/N) of the pulse and instrumental characteristics. Using the radiometer equation, the peak flux is given by
\begin{equation}
    S_{\rm{peak}} = \beta \frac{(S/N)_{\rm{peak}} T_{\rm{sys}}}{G\sqrt{n_p\Delta\nu t}},
\end{equation}
where $T_{\rm{sys}}$ is the system temperature, $\Delta\nu$ is bandwidth, $G$ is telescope gain (in K/Jy), $n_p$ is the number of polarizations, and $\beta$ is the quantization efficiency \citep{Lorimer:2004}. The system temperature is often approximated by
\begin{equation}
    T_{\rm{sys}} \approx T_{\rm{receiver}} + T_{\rm{sky}}
\end{equation}
where $T_{\rm{receiver}}$ is the receiver temperature, and $T_{\rm{sky}}$ is the contribution from the sky. Note that this approximation ignores several contributors to the noise budget, such as the telescope spillover and the atmosphere/ionosphere; \citep[see for example,][]{Campbell:2002}. 

In the pulsar and FRB literature, it remains common for the sky contribution to be estimated based on the \citet{Haslam:1982} all-sky map, with a power-law frequency scaling applied to account for the sky-averaged synchrotron spectrum:
\begin{equation}
    T_{\rm{sky}}(\nu,\theta,\phi) \approx (T_{\rm{408}}(\theta,\phi) - T_{\rm{CMB}}) \left(\frac{\nu}{408\rm{\,MHz}} \right)^{\beta} + T_{\rm{CMB}}.
\end{equation}
Here, $T_{408}$ is the temperature given by the Haslam map, $T_{\rm{CMB}}$ is the cosmic microwave background\footnote{The zero level of Haslam map includes the CMB contribution; however, the $T_{\rm{CMB}}$ term is often ignored when scaling.}, and $\beta$ is known as the spectral index. The Haslam map is a radio continuum all-sky map taken at 408 MHz, which combines data from Northern and Southern hemisphere surveys. This map remains in widespread usage;  a improved version with striping artefacts and point sources removed was provided by \citep{Remazeilles:2015}.

A primary issue of the spectral-index-scaled Haslam estimation (henceforth SISH) is that it ignores spectral variations across the sky, and frequency evolution of the spectral index. The SISH estimate was once a common approach among cosmologists, but is no longer used as better accuracy can be attained from models that incorporate sky maps across a range of frequencies. I argue that the SISH approach should also be deprecated within pulsar and FRB studies, in favor of using more detailed models of diffuse Galactic radio emission. 

\section{Global diffuse sky models}

The spectral index of Galactic synchrotron emission is known to vary with observing frequency and direction. For example, recent Southern-hemisphere observations across 50--100 MHz show a sky-averaged spectral index of $-2.59 < \beta < -2.54$, between 0 and 12 h LST, flattening to $\beta = -2.46 \pm 0.011$ at 18.2h \citep{Mozdzen:2019}. At higher frequencies the spectral index is shown to steepen; recent observations of the North celestial pole across 4.7--22.8\,GHz pole exhibit $\beta=-2.91 \pm 0.04$ \citep{Dickinson:2019}.

Several models of diffuse Galactic emission are publicly available. The Global Sky Model (GSM) of \citet{deOliveira-Costa:2008} is an all-sky model of diffuse Galactic radio emission across 10 MHz to 100 GHz. The GSM is generated from a principal component fit to 11 large-area radio surveys -- including \citet{Haslam:1982}. \citet{Zheng:2017} use a similar approach, but extend the principal component analysis approach to handle non-overlapping regions, enabling the inclusion of 29 sky maps across 10 MHz to 10 THz (henceforth GSM16).  \citet{Dowell:2017} provide a low-frequency sky model (LFSM, 10--408\,MHz) that incorporates new observations across 35--80\,MHz. Note that the Haslam 408 MHz is included in each sky model.

\section{A Python interface to Global sky models}

The interfaces to GSM, GSM16 and LFSM are written in Fortran, C, and Python, respectively. The {\sc{PyGDSM}} package\footnote{\url{https://github.com/telegraphic/pygdsm}} \citet{Price:2016} provides a unified Python interface to these sky models. The {\sc{PyGDSM}} interface to each model provides a method {\tt{get\_sky\_temperature()}}, which returns the sky temperature (in K) for a given pointing and frequency. For example:

\begin{minted}{python}
# Import packages
import numpy as np
from pygdsm import GlobalSkyModel
from astropy.coordinates import SkyCoord

# Setup observing frequency and pointing direction
freqs_mhz = np.array([100, 200, 300, 400])   # In MHz
sky_coord = SkyCoord("17h29m", "-2d7m", unit=('hourangle', 'degree'), frame='galactic')

# Generate T_sky from GSM
gsm = GlobalSkyModel()
T_sky = gsm.get_sky_temperature(sky_coord, freqs_mhz)

# Output T_sky = array([3280.90,  611.91,  223.22,  108.22]) (rounded)
\end{minted}

This code snippet shows that sky models can be easily used in lieu of the SISH approach. A skymap generated from {\sc{PyGDSM}} using the \citet{deOliveira-Costa:2008} GSM model is shown in Fig 1; as can be seen, $\beta$ varies across the sky, and is flatter within the Galactic plane (where a large fraction of pulsars reside). However, note that $\beta$ also varies as a function of frequency, so the map of Fig. 1 is not accurate outside 400--800\,MHz. 

\section{Use of frequency-scaled Haslam maps in FRB studies}

\begin{figure*}
    \centering
    \includegraphics[width=1.0\columnwidth]{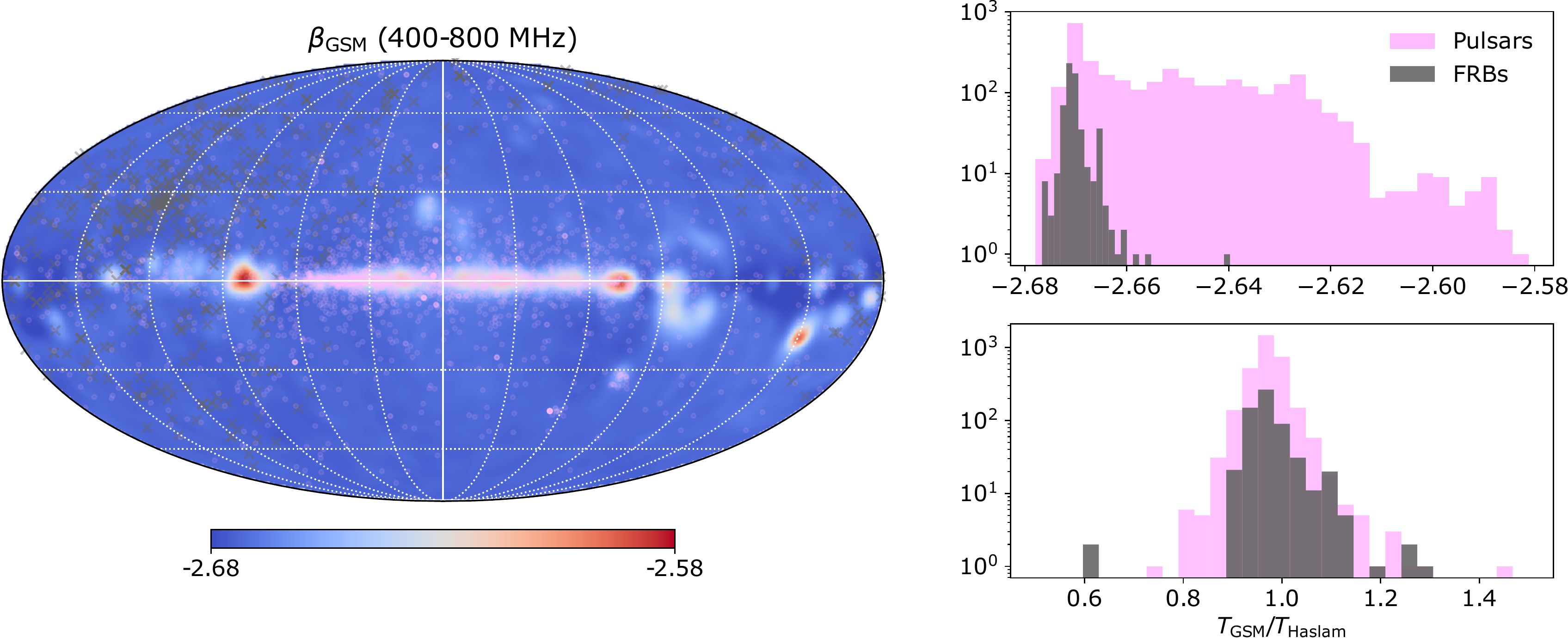}%
    \caption{(L) A map of spectral index between 400--800 MHz, computed using the GSM \citep{deOliveira-Costa:2008}, with locations of FRBs from the CHIME FRB catalog and ATNF pulsar database overlaid. (R) Histogram of computed spectral indices from the GSM map at FRB locations (top panel), and the corresponding ratio of $T_{\rm{sky}}$ estimate from the GSM to the SISH estimate (bottom panel). \label{fig:diagram}}
\end{figure*}

To show the prevalence of the SISH approach in recent transient literature, I searched the Astrophysics Data System (ADS\footnote{\url{https://ui.adsabs.harvard.edu}, accessed 28-Sep-2021, using the search term {\tt{citations(bibcode:1982A\&AS...47....1H)  pulsar}}.}) for papers including the term `pulsar' with a citation to \citet{Haslam:1982}. In total, 179 results were returned. Limiting the search to papers published in 2020, fifteen papers were returned by ADS, of which eleven employed SISH. Of these, seven papers used a value of $\beta=-2.55$ \citep{Bilous:2020,Bondonneau:2020,Champion:2020,Chawla:2020,vanLeeuwen:2020,Parent:2020,Tan:2020}, three papers used $\beta=-2.6$ \citep{Agarwal:2020, Basu:2020, Oostrum:2020}, and one paper used $\beta=-2.5$ \citep{Good:2020}.

To illustrate how the $T_{\rm{sky}}$ contribution can affect flux calibration, I generated $\beta$ estimates toward FRBs from the Canadian Hydrogen Intensity Mapping Experiment (CHIME) FRB catalog \citep{TheCHIME/FRBCollaboration:2021}, queried via the {\sc{Cfod}}\footnote{\url{https://github.com/chime-frb-open-data/chime-frb-open-data}} package, and the ATNF Pulsar database, queried by \textsc{psrqpy} package \citep{psrqpy}. Fig. 1 shows an all-sky map (Mollweide projection) of spectral index across 400--800 MHz (the CHIME observing band), with FRB and pulsar locations overlaid. The GSM predicts a median spectral index $\beta\approx-2.67$ across the CHIME band (Fig 1. top right panel). The corresponding ratio of GSM $T_{\rm{sky}}$  estimate against the SISH estimate is shown in the bottom panel. While most values are within $\sim\pm20\%$, there are some outliers. That said, the $T_{\rm{receiver}}$ for telescopes at these frequencies is greater than $T_{\rm{sky}}$ for most sight lines, so overall $T_{\rm{sys}}$ estimates will vary by a smaller percentage. 

As the sky is brighter at lower frequencies, there are larger magnitude differences between SISH and GSM-based estimates at frequencies below the Haslam 408 MHz map. Recently, the repeating FRB 20180916B was detected across 110--188 MHz \citep{Pleunis:2021}, the lowest-frequency FRB detection to date. At these frequencies the GSM predicts an average $\beta\approx-2.53$ off the Galactic plane ($|l| <$ 2) and $\beta\approx-2.43$ within 2 degrees of the plane. The GSM and SISH approaches differ by an average $\sim11\%$ toward the plane, and $\sim2\%$ off the plane.
 
\section{Concluding remarks}

In summary, SISH-based estimates of $T_{\rm{sys}}$ are likely to be accurate to within a few percent of GSM-based estimates across 110-800 MHz. Nevertheless, I advise against using SISH approach if 1) the source is within the Galactic plane, or 2) if the observing frequency is below $\sim$100 MHz or above $\sim$1.5 GHz (at which point other contributions to the GSM start to become apparent). As most pulsars are within the Galactic plane, I recommend against using the SISH approach to estimate $T_{\rm{sys}}$ in pulsar studies. 

Global diffuse sky models are specifically designed to provide more accurate estimates of $T_{\rm{sky}}$ than the SISH method, and as such should be preferred where possible. Note that estimates of $T_{\rm{sky}}$  can also be improved by taking into account the telescope's beam pattern and weighting the sky map appropriately, and/or accounting for second-order effects such as ionospheric and atmospheric noise contributions. Alternatively, flux estimates based on observations of calibrator sources, and/or instrument-specific calibration can be used instead. 

The {\sc{PyGDSM}} package provides an interface to the GSM, GSM16 and LFSM sky models, as well as the SISH method. Future sky models will be informed by recent, improved and upcoming all-sky maps of diffuse emission \citep[e.g.][]{Eastwood:2018,Jones:2018,Carretti:2019,Monsalve:2021}, which will further improve their accuracy against the SISH method.

Although SISH estimates may be `good enough' for current science goals, future requirements and new low-frequency and wide-bandwidth instruments motivate switching to better methods. As there is little impediment to using global diffuse sky models in FRB and pulsar data analysis, I suggest that the use of SISH estimates should be discouraged.

\section{Software}
This work made use of {\sc{Astropy}} \citep{AstropyCollaboration:2013}, {\sc{Numpy}} \citep{Harris:2020},  {\sc{matplotlib}} \citep{Hunter:2007}, {\sc{PyGDSM}} \citep{Price:2016}, the CHIME/FRB open data package ({\sc{Cfod}}), and {\sc{psrqpy}} \citep{psrqpy}.


\bibliographystyle{aasjournal}
\vspace{1 cm}
\bibliography{references}

\begin{thebibliography}{}
\expandafter\ifx\csname natexlab\endcsname\relax\def\natexlab#1{#1}\fi
\providecommand{\url}[1]{\href{#1}{#1}}

\bibitem[{{Agarwal} {et~al.}(2020){Agarwal}, {Lorimer}, {Surnis}, {Pei},
  {Karastergiou}, {Golpayegani}, {Werthimer}, {Cobb}, {McLaughlin}, {White},
  {Armour}, {MacMahon}, {Siemion}, \& {Foster}}]{Agarwal:2020}
{Agarwal}, D., {Lorimer}, D.~R., {Surnis}, M.~P., {et~al.} 2020, \mnras, 497,
  352

\bibitem[{{Astropy Collaboration} {et~al.}(2013){Astropy Collaboration},
  {Robitaille}, {Tollerud}, {Greenfield}, {Droettboom}, {Bray}, {Aldcroft},
  {Davis}, {Ginsburg}, {Price-Whelan}, {Kerzendorf}, {Conley}, {Crighton},
  {Barbary}, {Muna}, {Ferguson}, {Grollier}, {Parikh}, {Nair}, {Unther},
  {Deil}, {Woillez}, {Conseil}, {Kramer}, {Turner}, {Singer}, {Fox}, {Weaver},
  {Zabalza}, {Edwards}, {Azalee Bostroem}, {Burke}, {Casey}, {Crawford},
  {Dencheva}, {Ely}, {Jenness}, {Labrie}, {Lim}, {Pierfederici}, {Pontzen},
  {Ptak}, {Refsdal}, {Servillat}, \& {Streicher}}]{AstropyCollaboration:2013}
{Astropy Collaboration}, {Robitaille}, T.~P., {Tollerud}, E.~J., {et~al.} 2013,
  \aap, 558, A33

\bibitem[{{Basu} {et~al.}(2020){Basu}, {Joshi}, {Krishnakumar}, {Bhattacharya},
  {Nandi}, {Bandhopadhay}, {Char}, \& {Manoharan}}]{Basu:2020}
{Basu}, A., {Joshi}, B.~C., {Krishnakumar}, M.~A., {et~al.} 2020, \mnras, 491,
  3182

\bibitem[{{Bilous} {et~al.}(2020){Bilous}, {Bondonneau}, {Kondratiev},
  {Grie{\ss}meier}, {Theureau}, {Hessels}, {Kramer}, {van Leeuwen}, {Sobey},
  {Stappers}, {ter Veen}, \& {Weltevrede}}]{Bilous:2020}
{Bilous}, A.~V., {Bondonneau}, L., {Kondratiev}, V.~I., {et~al.} 2020, \aap,
  635, A75

\bibitem[{{Bondonneau} {et~al.}(2020){Bondonneau}, {Grie{\ss}meier},
  {Theureau}, {Bilous}, {Kondratiev}, {Serylak}, {Keith}, \&
  {Lyne}}]{Bondonneau:2020}
{Bondonneau}, L., {Grie{\ss}meier}, J.~M., {Theureau}, G., {et~al.} 2020, \aap,
  635, A76

\bibitem[{{Campbell}(2002)}]{Campbell:2002}
{Campbell}, D.~B. 2002, in Astronomical Society of the Pacific Conference
  Series, Vol. 278, Single-Dish Radio Astronomy: Techniques and Applications,
  ed. S.~{Stanimirovic}, D.~{Altschuler}, P.~{Goldsmith}, \& C.~{Salter},
  81--90

\bibitem[{{Carretti} {et~al.}(2019){Carretti}, {Haverkorn}, {Staveley-Smith},
  {Bernardi}, {Gaensler}, {Kesteven}, {Poppi}, {Brown}, {Crocker}, {Purcell},
  {Schnitzeler}, \& {Sun}}]{Carretti:2019}
{Carretti}, E., {Haverkorn}, M., {Staveley-Smith}, L., {et~al.} 2019, \mnras,
  489, 2330

\bibitem[{{Champion} {et~al.}(2020){Champion}, {Cognard}, {Cruces},
  {Desvignes}, {Jankowski}, {Karuppusamy}, {Keith}, {Kouveliotou}, {Kramer},
  {Liu}, {Lyne}, {Mickaliger}, {O'Connor}, {Parthasarathy}, {Porayko},
  {Rajwade}, {Stappers}, {Torne}, {van der Horst}, \&
  {Weltevrede}}]{Champion:2020}
{Champion}, D., {Cognard}, I., {Cruces}, M., {et~al.} 2020, \mnras, 498, 6044

\bibitem[{{Chawla} {et~al.}(2020){Chawla}, {Andersen}, {Bhardwaj}, {Fonseca},
  {Josephy}, {Kaspi}, {Michilli}, {Pleunis}, {Bandura}, {Bassa}, {Boyle},
  {Brar}, {Cassanelli}, {Cubranic}, {Dobbs}, {Dong}, {Gaensler}, {Good},
  {Hessels}, {Landecker}, {Leung}, {Li}, {Lin}, {Masui}, {Mckinven},
  {Mena-Parra}, {Merryfield}, {Meyers}, {Naidu}, {Ng}, {Patel},
  {Rafiei-Ravandi}, {Rahman}, {Sanghavi}, {Scholz}, {Shin}, {Smith}, {Stairs},
  {Tendulkar}, \& {Vanderlinde}}]{Chawla:2020}
{Chawla}, P., {Andersen}, B.~C., {Bhardwaj}, M., {et~al.} 2020, \apjl, 896, L41

\bibitem[{{de Oliveira-Costa} {et~al.}(2008){de Oliveira-Costa}, {Tegmark},
  {Gaensler}, {Jonas}, {Landecker}, \& {Reich}}]{deOliveira-Costa:2008}
{de Oliveira-Costa}, A., {Tegmark}, M., {Gaensler}, B.~M., {et~al.} 2008,
  \mnras, 388, 247

\bibitem[{{Dickinson} {et~al.}(2019){Dickinson}, {Barr}, {Chiang}, {Copley},
  {Grumitt}, {Harper}, {Heilgendorff}, {Jew}, {Jonas}, {Jones}, {Leahy},
  {Leech}, {Leitch}, {Muchovej}, {Pearson}, {Peel}, {Readhead}, {Sievers},
  {Stevenson}, \& {Taylor}}]{Dickinson:2019}
{Dickinson}, C., {Barr}, A., {Chiang}, H.~C., {et~al.} 2019, \mnras, 485, 2844

\bibitem[{{Dowell} {et~al.}(2017){Dowell}, {Taylor}, {Schinzel}, {Kassim}, \&
  {Stovall}}]{Dowell:2017}
{Dowell}, J., {Taylor}, G.~B., {Schinzel}, F.~K., {Kassim}, N.~E., \&
  {Stovall}, K. 2017, \mnras, 469, 4537

\bibitem[{{Eastwood} {et~al.}(2018){Eastwood}, {Anderson}, {Monroe},
  {Hallinan}, {Barsdell}, {Bourke}, {Clark}, {Ellingson}, {Dowell}, {Garsden},
  {Greenhill}, {Hartman}, {Kocz}, {Lazio}, {Price}, {Schinzel}, {Taylor},
  {Vedantham}, {Wang}, \& {Woody}}]{Eastwood:2018}
{Eastwood}, M.~W., {Anderson}, M.~M., {Monroe}, R.~M., {et~al.} 2018, \aj, 156,
  32

\bibitem[{{Good} {et~al.}(2020){Good}, {Andersen}, {Chawla}, {Crowter}, {Dong},
  {Fonseca}, {Meyers}, {Ng}, {Pleunis}, {Ransom}, {Stairs}, {Tan}, {Bhardwaj},
  {Boyle}, {Dobbs}, {Gaensler}, {Kaspi}, {Masui}, {Naidu}, {Rafiei-Ravandi},
  {Scholz}, {Smith}, \& {Tendulkar}}]{Good:2020}
{Good}, D.~C., {Andersen}, B.~C., {Chawla}, P., {et~al.} 2020, arXiv e-prints,
  arXiv:2012.02320

\bibitem[{{Harris} {et~al.}(2020){Harris}, {Millman}, {van der Walt},
  {Gommers}, {Virtanen}, {Cournapeau}, {Wieser}, {Taylor}, {Berg}, {Smith},
  {Kern}, {Picus}, {Hoyer}, {van Kerkwijk}, {Brett}, {Haldane}, {del R{\'\i}o},
  {Wiebe}, {Peterson}, {G{\'e}rard-Marchant}, {Sheppard}, {Reddy}, {Weckesser},
  {Abbasi}, {Gohlke}, \& {Oliphant}}]{Harris:2020}
{Harris}, C.~R., {Millman}, K.~J., {van der Walt}, S.~J., {et~al.} 2020, \nat,
  585, 357

\bibitem[{{Haslam} {et~al.}(1982){Haslam}, {Salter}, {Stoffel}, \&
  {Wilson}}]{Haslam:1982}
{Haslam}, C.~G.~T., {Salter}, C.~J., {Stoffel}, H., \& {Wilson}, W.~E. 1982,
  \aaps, 47, 1

\bibitem[{{Hunter}(2007)}]{Hunter:2007}
{Hunter}, J.~D. 2007, Computing in Science and Engineering, 9, 90

\bibitem[{{Jones} {et~al.}(2018){Jones}, {Taylor}, {Aich}, {Copley}, {Chiang},
  {Davis}, {Dickinson}, {Grumitt}, {Hafez}, {Heilgendorff}, {Holler}, {Irfan},
  {Jew}, {John}, {Jonas}, {King}, {Leahy}, {Leech}, {Leitch}, {Muchovej},
  {Pearson}, {Peel}, {Readhead}, {Sievers}, {Stevenson}, \&
  {Zuntz}}]{Jones:2018}
{Jones}, M.~E., {Taylor}, A.~C., {Aich}, M., {et~al.} 2018, \mnras, 480, 3224

\bibitem[{{Lorimer} \& {Kramer}(2004)}]{Lorimer:2004}
{Lorimer}, D.~R., \& {Kramer}, M. 2004, {Handbook of Pulsar Astronomy}, Vol.~4

\bibitem[{{Monsalve} {et~al.}(2021){Monsalve}, {Rogers}, {Bowman}, {Mahesh},
  {Murray}, {Mozdzen}, {Johnson}, {Barrett}, {Samson}, \&
  {Lewis}}]{Monsalve:2021}
{Monsalve}, R.~A., {Rogers}, A. E.~E., {Bowman}, J.~D., {et~al.} 2021, \apj,
  908, 145

\bibitem[{{Mozdzen} {et~al.}(2019){Mozdzen}, {Mahesh}, {Monsalve}, {Rogers}, \&
  {Bowman}}]{Mozdzen:2019}
{Mozdzen}, T.~J., {Mahesh}, N., {Monsalve}, R.~A., {Rogers}, A.~E.~E., \&
  {Bowman}, J.~D. 2019, \mnras, 483, 4411

\bibitem[{{Oostrum} {et~al.}(2020){Oostrum}, {van Leeuwen}, {Maan}, {Coenen},
  \& {Ishwara-Chandra}}]{Oostrum:2020}
{Oostrum}, L.~C., {van Leeuwen}, J., {Maan}, Y., {Coenen}, T., \&
  {Ishwara-Chandra}, C.~H. 2020, \mnras, 492, 4825

\bibitem[{{Parent} {et~al.}(2020){Parent}, {Chawla}, {Kaspi}, {Agazie},
  {Blumer}, {DeCesar}, {Fiore}, {Fonseca}, {Hessels}, {Kaplan}, {Kondratiev},
  {LaRose}, {Levin}, {Lewis}, {Lynch}, {McEwen}, {McLaughlin}, {Mingyar}, {Al
  Noori}, {Ransom}, {Roberts}, {Schmiedekamp}, {Schmiedekamp}, {Siemens},
  {Spiewak}, {Stairs}, {Surnis}, {Swiggum}, \& {van Leeuwen}}]{Parent:2020}
{Parent}, E., {Chawla}, P., {Kaspi}, V.~M., {et~al.} 2020, \apj, 904, 92

\bibitem[{{Pitkin}(2018)}]{psrqpy}
{Pitkin}, M. 2018, {Journal of Open Source Software}, 3, 538.
\newblock \url{https://doi.org/10.21105/joss.00538}

\bibitem[{{Pleunis} {et~al.}(2021){Pleunis}, {Michilli}, {Bassa}, {Hessels},
  {Naidu}, {Andersen}, {Chawla}, {Fonseca}, {Gopinath}, {Kaspi}, {Kondratiev},
  {Li}, {Bhardwaj}, {Boyle}, {Brar}, {Cassanelli}, {Gupta}, {Josephy},
  {Karuppusamy}, {Keimpema}, {Kirsten}, {Leung}, {Marcote}, {Masui},
  {Mckinven}, {Meyers}, {Ng}, {Nimmo}, {Paragi}, {Rahman}, {Scholz}, {Shin},
  {Smith}, {Stairs}, \& {Tendulkar}}]{Pleunis:2021}
{Pleunis}, Z., {Michilli}, D., {Bassa}, C.~G., {et~al.} 2021, \apjl, 911, L3

\bibitem[{{Price}(2016)}]{Price:2016}
{Price}, D.~C. 2016, {PyGSM: Python interface to the Global Sky Model}, , ,
  ascl:1603.013

\bibitem[{{Remazeilles} {et~al.}(2015){Remazeilles}, {Dickinson}, {Banday},
  {Bigot-Sazy}, \& {Ghosh}}]{Remazeilles:2015}
{Remazeilles}, M., {Dickinson}, C., {Banday}, A.~J., {Bigot-Sazy}, M.~A., \&
  {Ghosh}, T. 2015, \mnras, 451, 4311

\bibitem[{{Tan} {et~al.}(2020){Tan}, {Bassa}, {Cooper}, {Hessels},
  {Kondratiev}, {Michilli}, {Sanidas}, {Stappers}, {van Leeuwen}, {Donner},
  {Grie{\ss}meier}, {Kramer}, {Tiburzi}, {Weltevrede}, {Ciardi}, {Hoeft},
  {Mann}, {Miskolczi}, {Schwarz}, {Vocks}, \& {Wucknitz}}]{Tan:2020}
{Tan}, C.~M., {Bassa}, C.~G., {Cooper}, S., {et~al.} 2020, \mnras, 492, 5878

\bibitem[{{The CHIME/FRB Collaboration} {et~al.}(2021){The CHIME/FRB
  Collaboration}, {:}, {Amiri}, {Andersen}, {Bandura}, {Berger}, {Bhardwaj},
  {Boyce}, {Boyle}, {Brar}, {Breitman}, {Cassanelli}, {Chawla}, {Chen},
  {Cliche}, {Cook}, {Cubranic}, {Curtin}, {Deng}, {Dobbs}, {Fengqiu}, {Dong},
  {Eadie}, {Fandino}, {Fonseca}, {Gaensler}, {Giri}, {Good}, {Halpern}, {Hill},
  {Hinshaw}, {Josephy}, {Kaczmarek}, {Kader}, {Kania}, {Kaspi}, {Landecker},
  {Lang}, {Leung}, {Li}, {Lin}, {Masui}, {Mckinven}, {Mena-Parra},
  {Merryfield}, {Meyers}, {Michilli}, {Milutinovic}, {Mirhosseini},
  {M{\"u}nchmeyer}, {Naidu}, {Newburgh}, {Ng}, {Patel}, {Pen}, {Petroff},
  {Pinsonneault-Marotte}, {Pleunis}, {Rafiei-Ravandi}, {Rahman}, {Ransom},
  {Renard}, {Sanghavi}, {Scholz}, {Shaw}, {Shin}, {Siegel}, {Sikora}, {Singh},
  {Smith}, {Stairs}, {Tan}, {Tendulkar}, {Vanderlinde}, {Wang}, {Wulf}, \&
  {Zwaniga}}]{TheCHIME/FRBCollaboration:2021}
{The CHIME/FRB Collaboration}, {:}, {Amiri}, M., {et~al.} 2021, arXiv e-prints,
  arXiv:2106.04352

\bibitem[{{van Leeuwen} {et~al.}(2020){van Leeuwen}, {Mikhailov}, {Keane},
  {Coenen}, {Connor}, {Kondratiev}, {Michilli}, \& {Sanidas}}]{vanLeeuwen:2020}
{van Leeuwen}, J., {Mikhailov}, K., {Keane}, E., {et~al.} 2020, \aap, 634, A3

\bibitem[{{Zheng} {et~al.}(2017){Zheng}, {Tegmark}, {Dillon}, {Kim}, {Liu},
  {Neben}, {Jonas}, {Reich}, \& {Reich}}]{Zheng:2017}
{Zheng}, H., {Tegmark}, M., {Dillon}, J.~S., {et~al.} 2017, \mnras, 464, 3486

\end{thebibliography}

\end{document}